\documentclass[aps,showpacs,twocolumn]{revtex4-1}

\usepackage{graphicx,epsf}
\usepackage{float}
\usepackage{color}
\usepackage{amsfonts,amsmath,amssymb}
\usepackage{stmaryrd,wasysym}
\usepackage{ulem}

\newcommand{\red}{\textcolor{black}}

\begin{document}

\title{Snaking states on a cylindrical surface in a perpendicular 
magnetic field}

\author{Andrei Manolescu,$^1$ Tomas Orn Rosdahl,$^2$ Sigurdur Erlingsson,$^1$ Llorens Serra,$^3$ Vidar Gudmundsson$^2$}

\affiliation{$^1$School of Science and Engineering, Reykjavik University,
Menntavegur 1, IS-101 Reykjavik, Iceland,\\
$^2$Science Institute, University of Iceland, Dunhaga 3, IS-107 Reykjavik, Iceland, \\
$^3$IFISC (CSIC-UIB) and Department of Physics, University of the Balearic Islands,
E-07122 Palma de Mallorca, Spain}

\begin{abstract}
We calculate electronic states on a closed cylindrical surface as a model
of a core-shell nanowire. The length of the cylinder can be infinite or
finite.  We define cardinal points on the circumference of the cylinder
and consider a spatially uniform magnetic field perpendicular to the
cylinder axis, in the direction South-North. The orbital motion of
the electrons depends on the radial component of the field which is 
nonuniform around the circumference: it is equal to the total field at North
and South, but vanishes at the West and East sides.  For a strong field,
when the magnetic length is comparable to the radius of the cylinder,
the electronic states at North and South become localized cyclotron
orbits, whereas at East and West the states become long and narrow snaking
orbits propagating along the cylinder.  The energy of the cyclotron states
increases with the magnetic field whereas the energy of the snaking states
is stable.  Consequently, at high magnetic fields the electron density
vanishes at North and South and concentrates at East and West.  We include
spin-orbit interaction with linear Rashba and Dresselhaus models.
For a cylinder of finite length the Dresselhaus interaction produces
an axial twist of the charge density relative to the center of the wire,
which may be amplified in the presence of the Rashba interaction.
\end{abstract}



\maketitle

\section{Introduction}
\label{Intro}

In the beginning of the 90's nanomagnets of 30-100 nm diameter were
fabricated using a scanning tunneling microscope \cite{McCord1990}.
Soon after that electronic transport in the presence of nonuniform
magnetic fields created by magnetic steps, magnetic barriers, or magnetic
wells, have been studied and interesting conductance properties have
been predicted \cite{Peeters1993}.  For example, the tunneling
probability of an electron through a magnetic barrier which depends not only
on the momentum component perpendicular to the barrier (like for an
electric barrier), but also on the component parallel to the barrier.
The nonuniform magnetic field is formally equivalent to an effective
potential which depends on the momentum \cite{Matulis1994}.
This is in fact the idea of magnetic lenses used to focus the 
electron beam, for example in cathode tubes.
\red{In nanostructured magnetic barriers additional scattering or 
quantization effects could be experimentally detected 
\cite{Cerchez2007,Tarasov2010}.}

Then, periodic magnetic fields over hundreds of nm were created in
GaAs heterostructures.  One design was with superconducting stripes
placed on the surface of the semiconductor heterostructure, but which
produced only weak variations of the magnetic field at the level of
the two-dimensional electron gas \cite{Carmona1995}.  Another design was
with ferromagnetic stripes grown at the surface, which produced fields
with much stronger gradients, and even changing sign \cite{Ye1995}.
In some regions the field component perpendicular to the two-dimensional
electron system was ``up'' and in other regions it was ``down''.  In such
a situation, for an appropriate orientation of the electron velocity along
the zero-field line, the Lorentz forces on both sides tend to bend the
trajectory back, towards the line, such that the electron remains trapped
in an open snaking orbit \cite{Muller1992}.  For this reason a strong
positive magnetoresistance was observed in the experiments \cite{Ye1995}.
For a large amplitude of the periodic magnetic field (like half a tesla)
the electronic states were a superposition of two dimensional cyclotron
orbits and one-dimensional free electron states along the snaking 
paths \cite{Ibrahim1995,Zwerschke1999}.

Another interesting example of electrons in an inhomogeneous magnetic
field is when the electrons are situated on a cylindrical surface.
Such an electron system has been built by several groups using the strain
architecture technology \cite{Cho2006}.  By etching at a certain depth
underneath an InGaAs heterostructure the upper layers automatically
bend, due to the strain relaxation, until becoming a cylinder with
radius of 10-20 $\mu{\rm m}$ \cite{Friedland2007,Friedland2009}.  
The curved surface is never closed, but it can also be  
rolled up many times until it looks like a carpet roll \cite{Gayer2012}.
In the presence of a magnetic field perpendicular to the axis of the 
cylinder the Lorentz force depends only on the radial component and
vanishes at right angles with respect to the external field. 

Four-terminal transport measurements on a small region defined on such
a curved surface, with the zero-field line inside the transport
region, show oscillations of the magnetoresistance indicating that
the electrons scatter into the snaking orbits and then deviate either
towards the positive side or towards the negative side of the magnetic
field.  Although the two-dimensional electron gas has high mobility,
it is not clear whether the electronic transport is truly ballistic
in this case due to the relatively rough boundaries of the transport
region. An interesting hypothesis  is that the Dresselhaus spin-orbit
effect is responsible for the preferential deviation of the 
electrons \cite{Friedland2008}.

With a totally different technology, thin nanowires of diameters down
to 100 nm or less can now be grown \cite{Thelander2006}.  Field effect
transistors with high electron mobility have been realized with InAs/InP
core-shell cylinders \cite{Bryllert2006}.  
\red {In that case the InAs core was conductor while the InP shell
was insulating, with a passivation role.  Tubular conductors
have also been created.  For example with nanowires made of InN
\cite{Richter2008,Blomers2008} or InAs \cite{Wirths2011}, where
the tubular conductor is formed due to the accumulation layer at the
surface resulting from the Fermi level pinning.  More recently GaAs/InAs
core-shell structures have been made, where now an insulating core of GaAs is surrounded
by a conductive layer of InAs \cite{Blomers2011}.}
Transport properties of phase-coherent type in GaAs/InAs
core-shell nanowires with a magnetic field in different directions, and
also temperature dependent measurements, have been recently reported
\cite{Blomers2011,Blomers2013}. Conductance oscillations are observed
as functions of a longitudinal magnetic field and a gate voltage, and
also an antilocalization effect due to the electron spin.  Hall effect
measurements are performed on InAs nanowires \cite{Blomers2012}.  However,
to our knowledge, systematic magnetotransport measurements  with the
magnetic field perpendicular to the cylinder are not reported, and the
role or the presence of the snaking orbits is not clear.


Several theoretical papers addressed the quantum states and some
transport properties of electrons situated on a closed cylindrical
surface in the presence of an external magnetic field.  Oscillations of
magnetoconductance in the case of ballistic transport were described
as interference effects between electrons propagating through different
channels along the cylinder \cite{Tserkovnyak2006}.  Oscillations of
Aharonov-Bohm type are also predicted for the magnetizations of core-shell
nanowires \cite{Gladilin2013}.  In particular, an interesting paper by
Ferrari et al. \cite{Ferrari2008} describes states in a transverse magnetic
field, possibly periodic along the length of the cylinder.  The authors
give a hint of the snaking orbits, but they do not focus on them. In more
recent papers from the same group \cite{Ferrari2009a,Ferrari2009b,Royo2013} 
core-shell nanowires with hexagonal cross section are
considered, which is a more realistic shape for narrow cylinders
\cite{Wong2011,Rieger2012,Haas2013}.  A detailed semi-classical analysis
of cyclotron and snaking states on the cylinder in a transverse magnetic
field was made by Bellucci and Onorato \cite{Bellucci2010}.

Most of the previous theoretical studies of the cylindrical electron
gas in a transversal magnetic field analyzed only individual states,
cyclotronic, or snaking, or a combination of them, but not the
distribution of electrons on the cylinder surface when many such states
are combined.  For example, it is not clear under what conditions the
snaking states have a prominent presence in the electron density (and
consequently in transport) and when they are washed out by the other states.
In this paper we reconsider the quantum mechanics of electrons on
a cylindrical surface in a perpendicular magnetic field.  We show that
for a sufficiently strong, but realistic  magnetic field, perpendicular
to the axis of the cylinder, the electron density concentrates on the
lateral sides of the cylinder along the snaking orbits, whereas the
top and the bottom regions become totally depleted.  This result may
possibly be guessed from previous theoretical papers, but it has not been
explicitly shown.  We include systematically the spin and spin-orbit
interaction (SOI) within standard models incorporating a Rashba effect
\cite{Bringer2011} and a Dresselhaus contribution \cite{Friedland2008}
and we calculate the spin polarization around the surface for cylinders of
infinite and finite length.  In the latter case we obtain the distribution
of permanent currents closing over the edges of the cylinder.  
Inclusion of the Dresselhaus term produces a torsion effect wherein the charge
density is symmetric under a reflection over the $x$ or $y$ axes, simultaneous 
with the inversion of the $z$ coordinate.  This effect is related to the 
edge states.  

In Sec.\ \ref{QMM} we describe the model and we build the quantum mechanical states.
In Sec.\ \ref{SNO} we discuss the cyclotron and the snaking states.
In Sections\ \ref{NRI} and \ref{NRF} we show our results for cylinders of an infinite
and finite length, respectively.  Finally in Sec.\ \ref{CON} we summarize the 
conclusions.

\section{Quantum mechanical model for the infinite cylinder} \label{QMM}

\begin{figure} 
\begin{center}
\includegraphics [width=7 cm] {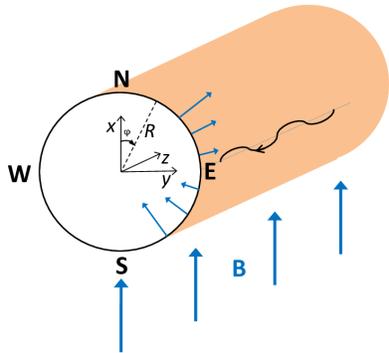}
\end{center}
\caption{An infinitely long cylinder in a perpendicular magnetic field
$B$ shown with the big arrows. The small arrows indicate the radial
field component on the cylinder surface.  Electrons are situated on the
cylinder surface and their orbital motion is determined by the radial
field component shown by the small arrows.  The ``cardinal'' points
(N,S,E,W) are indicated and used suggestively.  The radial field vanishes
at E and W points, i.\ e.\ along the dotted blue line where the electron
trajectory is an open snaking orbits in this region as shown by the
black wavy line.  At the N and S poles the orbits correspond to closed
cyclotron loops (not shown).  }
\label{Sample}
\end{figure}

We consider electrons on a cylindrical surface of zero thickness, radius
$R$, and infinite length, which is illustrated in Fig.\ \ref{Sample}. The
length of the cylinder is in the $z$ direction, the vertical axis is
$x$, and the horizontal axis is $y$.  For convenience we also name the
``cardinal'' points as North (N), South (S), East (E), and West (W).
In order to calculate the quantum mechanical states of 
electrons situated on this surface we describe the corresponding Hilbert 
space with the basis set 
$\langle\varphi z |mks\rangle \equiv \langle\varphi|m\rangle \langle z|k\rangle |s\rangle$, 
where
\begin{equation}
\langle\varphi|m\rangle = \frac{1}{\sqrt{2\pi}}e^{im\varphi}, 
\langle z|k\rangle = \frac{1}{\sqrt{L}}e^{ikz},
|s\rangle = |\pm 1\rangle. 
\label{basis}
\end{equation}
In this basis set the quantum number $m=0,\pm 1,\pm 2, ...$ describe the 
$z$-projection of the angular momentum, $k$ is the wave
vector in the $z$ direction and $L$ is the unspecified length of the cylinder.
The spin states are associated to an abstract binary vector $|s\rangle$,
conventionally chosen as eigenvectors of the spin angular momentum in
the same direction $z$.

The basis set are the eigenvectors of the Hamiltonian without magnetic field 
and spin,
\begin{equation}
H_0=-\frac{\hbar^2}{2m_{\rm eff} R^2}\frac{\partial^2}{\partial \varphi^2}
+\frac{{p_z}^2}{2m_{\rm eff}}\ ,
\label{H0}
\end{equation}
with corresponding eigenenergies
\begin{equation}
\varepsilon_{mk}=\frac{\hbar^2}{2m_{\rm eff} R^2} \left[ m^2 +(kR)^2\right].
\label{E0}
\end{equation}
The total single particle Hamiltonian of the problem, $H$, 
is given by the sum of $H_0$ above and contributions from the magnetic 
field, vector potential, and SOI shown below.

In the presence of a magnetic field along the $x$ direction,
${\bf B}=(B,0,0)$, the vector potential can be chosen as ${\bf
A}=(0,0,By)=(0,0,BR\sin\varphi)$. With the momentum transformation
$p_z \to p_z+eA_z=p_z+eBR\sin\varphi$ we obtain the orbital contribution
of the magnetic field,
\begin{equation}
H_{B}=\frac{eBR}{m_{\rm eff}}p_z\sin\varphi+\frac{(eBR)^2}{2 m_{\rm eff}}(\sin\varphi)^2.
\label{HB}
\end{equation}

The spin Zeeman term is then given by
\begin{equation}
H_Z=-\frac{1}{2} g_{\rm eff}\mu_B B\sigma_x,
\label{HZ}
\end{equation}
where the Zeeman energy is included through the effective $g$-factor
($g_{\rm eff}$) of the electron in the semiconductor host material and
where $\mu_B$ is Bohr's magneton (defined with the mass of the free
electron and the absolute value of the electron charge).

The spin-orbit interaction is described with the linear Rashba 
($H_R$) and Dresselhaus ($H_D$) models as follows.
\begin{equation}
H_R=\frac{\alpha}{\hbar}\left[\sigma_{\varphi}(p_z+eBR\sin\varphi)
-\sigma_z p_{\varphi}\right],
\label{HR}
\end{equation}
which corresponds to the radial confinement of the electrons on the cylindrical 
surface analogous to the planar confinement of the two-dimensional electron gas 
in an asymmetric quantum well \cite{Bringer2011}.  
Here $\sigma_{\varphi}=\sigma_y\cos{\varphi}-\sigma_x\sin{\varphi}$
and $p_{\varphi}=-i\hbar/R\partial_{\varphi}$ are the azimuthal 
Pauli matrix and momentum, respectively.

The second type of SOI is the Dresselhaus coupling in zincblende structures, 
due to the bulk inversion asymmetry, which is of the form
\begin{equation}
H_D= \frac{\gamma}{\hbar} \bigl ( p_x (p_y^2-p_z^2) \sigma_x+ \mbox{cyclic permutations} \bigr ),
\end{equation}
where $\gamma$ is a material parameter that determines the strength of
the spin orbit coupling \cite{Winkler}.  In a similar way as in deriving
Eq.\ (\ref{HR}) we assume that the electronic system contained in the shell
behaves like a two-dimensional electron gas wrapped around the core.  
Projecting down to the lower radial mode gives rise to
\begin{equation}
H_D=\frac{\beta}{\hbar}\left[\frac{1}{2}
(\sigma_{\varphi}p_{\varphi}+p_{\varphi}\sigma_{\varphi})-\sigma_z
(p_z+eBR\sin\varphi)\right],
\label{HD}
\end{equation}
which is the standard model for the two-dimensional semiconductor
system \cite{Winkler,Ihn}, but here symmetrized in the azimuthal direction
$\hat\varphi$.  Nanowires grown from materials that are zincblende in
the bulk can occur with a wurtzite configuration due to the structural
changes in the lattice.  
A nanowire with a wurtzite lattice has a linear SOI term \cite{Fasth2007}
which is different than for a zincblende.
Nevertheless, here we consider nanowires that still have a zincblende
structure, which is possible since the crystallographic phase can be
controlled at growth \cite{Joyce2010}.


The total Hamiltonian is now
\begin{equation}
H=H_0+H_B+H_Z+H_R+H_D ,
\label{H}
\end{equation}
whose eigenstates $|ak\rangle$ are expanded in the basis (\ref{basis}): 
\begin{eqnarray}
H| ak\rangle  &=& E_{ak} |ak \rangle, \nonumber \\
|ak\rangle &=& \sum_{ms} c_{a,ms}(k) |mks\rangle \ 
\equiv \sum_{q} c_{aq}(k) |qk\rangle.
\label{expansion}
\end{eqnarray}
The wave vector $k$ is a good quantum number and it is therefore
conserved.  For any fixed $k$ the label $a=1,2,3,...$ denotes
the eigenstates in the increasing order of the energies $E_{ak}$
and incorporates the mixed orbital ($m$) and spin ($s$) states.
The spin is not conserved due to the presence of SOI. For
simplicity a composite discrete label $q=\{m,s\}$ has been introduced.
The coefficients $c_{aq}(k)$ are found by a numerical diagonalization of
the matrix elements of $H$ calculated in the basis (\ref{basis}), for
independent values of $k$ on a discrete, but sufficiently dense grid,
with 200-400 points.  In the numerical calculations, for each $k$ value
the basis is truncated to $|m|\leq 50$.  All the matrix elements of the
Hamiltonian are straightforward integrals of trigonometric functions
which are calculated analytically and shown in the Appendix A.

Once the single particle states are known the chemical potential $\mu$ can be
calculated by fixing the number of electrons $N$ or the average electron 
density $\langle n \rangle$, and by solving the equation for the occupation 
numbers,
\begin{equation}
\langle n \rangle \equiv \frac {N}{2\pi RL}= \frac{1}{2\pi RL}\sum_{ak} 
{\cal F}_{ak}\, ,
\label{chempot}
\end{equation}
where ${\cal F}_{ak}={\cal F}[(E_{ak}-\mu)/k_BT]$ is the Fermi function.
For the infinite cylinder the summation over the wave vectors
$k$ is in fact $(1/2\pi)\int dk$ which is carried out numerically
on the grid.  The particle and spin densities, $n(\varphi)$ and
$s_{\alpha}(\varphi)$, ($\alpha=x,y,z$ denoting the spatial direction), 
are further calculated as 
\begin{eqnarray}
n(\varphi) &=& \sum_{ak}{\cal F}_{ak} \psi_{ak}^{\dagger}\psi_{ak}^{}(\varphi,z), \nonumber\\
\sigma_\alpha (\varphi) &=& \sum_{ak}{\cal F}_{ak} \psi_{ak}^{\dagger} 
\frac{\hbar}{2}\sigma_\alpha \psi_{ak}^{}(\varphi,z),
\label{edens}
\end{eqnarray}
where $\psi_{ak}(\varphi,z)\equiv \langle {\bf r} | ak
\rangle$ are the wave functions (spinors) in the position representation,
the scalar product in the spin space being implicitly assumed in both
versions of Eq.\ (\ref{edens}).  In the numerical implementation of the
particle and spin densities the wave functions are expanded in the basis
(\ref{basis}) and the $k$ integrals are stored in the matrix
\begin{equation}
G_{q_1q_2}=\frac{1}{2\pi}\sum_{a}\int dk
{\cal F}_{ak}c^*_{aq_1}(k)c^{}_{aq_2}(k).
\end{equation}
With the help of this matrix Eq.\ (\ref{edens}) can be written as
\begin{eqnarray}
n(\varphi) &=& \frac{1}{2\pi R}\sum_{q_1 q_2} e^{i(m_2-m_1)\varphi}\ G_{q_1 q_2}\ \delta_{s_1s_2}, \nonumber\\ 
s_{\alpha}(\varphi) &=& \frac{1}{2\pi R}\sum_{q_1 q_2} e^{i(m_2-m_1)\varphi}\ G_{q_1 q_2}\ 
\frac{\hbar}{2}(\sigma_{\alpha})_{s_1 s_2},
\label{csd}
\end{eqnarray}
where $q_1=\{m_1,s_1\}$ and $q_2=\{m_2,s_2\}$.

\section{Snaking orbits and the effective potential} \label{SNO}

In our setup the orbital motion of the electrons is determined by the
component of the magnetic field perpendicular to the surface, if we
neglect the additional (small) effects due to the SOI.  The electrons
experience a nonuniform radial magnetic field at different points on
the cylinder which vanishes at right angles relative to the external
field, i.\ e.\ at the W and E points, as shown in Fig.\ \ref{Sample}.
In terms of classical mechanics the trajectory of an electron along the
zero-field line is a snaking orbit \cite{Bellucci2010}. It is laterally
confined by the Lorentz force which from both sides of the zero-field
line tends to bend it back towards the line \cite{Muller1992}.
Only one orientation of the velocity is compatible with a snaking state.
In the regions where the radial field does not vanish the classical
trajectory of the electrons correspond to closed cyclotron orbits,
possibly distorted by the field gradient.  For a weak magnetic field or
for a cylinder with a small radius $R$ the two types of states (snaking
and cyclotron) cannot be really distinguished.  But for a strong field,
such that the magnetic length $\ell_B\equiv \sqrt{\hbar/eB} < R$ the two
type a states may become spatially separated.  The cyclotron orbits may
occupy the N and S regions (and the surroundings), whereas the snaking
states may concentrate around the W and E regions.

\begin{figure} 
\begin{center}
\includegraphics [width=7 cm] {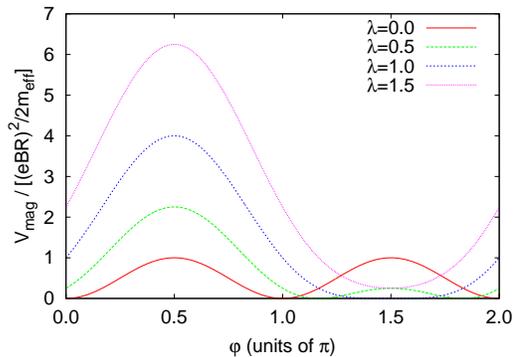}
\end{center}
\vspace{-5mm}
\caption{ The effective potential (\ref{Vmag}) around the cylinder
for different values of the dimensionless parameter $\lambda=k\ell_B^2/R$, 
vs. the polar angle $\varphi$.  
Electron states with low energies are bound to the minima of this
potential.  }
\label{Vmagfig}
\end{figure}

The confined snaking states can be easily explained by isolating
an effective ``magnetic" potential in the orbital Hamiltonian
\cite{Matulis1994,Ibrahim1995,Zwerschke1999,Ferrari2008}.  Since the
momentum in the $z$ direction $p_z=\hbar k$ is conserved we can look at
the matrix elements of the orbital terms of the Hamiltonian $H_O\equiv
H_0+H_B$ and write them in the form
\begin{equation}
H_O(k) =-\frac{\hbar^2}{2m_{\rm eff} R^2}\frac{\partial^2}{\partial \varphi^2}
+V_{\rm mag}(k,\varphi),
\nonumber
\end{equation}
where
\begin{equation}
V_{\rm mag}(k,\varphi)=\frac{(eBR)^2}{2 m_{\rm eff}}\left(\frac{k\ell_B^2}{R} + \sin \varphi\right)^2
\label{Vmag}
\end{equation}
is an effective potential which depends on the wave vector, or in fact
on the dimensionless parameter $\lambda=k\ell_B^2/R$, which can bind
(or localize) states with low energies 

For $|\lambda|<1$ the potential has
two minima, as seen in Fig.\ \ref{Vmagfig}, and hence the probability
densities are expected to show two maxima.  In particular, for $\lambda=0$
(or $k=0$), the states are confined on at $\varphi=0$ and $\varphi=\pi$,
i.\ e.\ at the N and S points.  For a $|\lambda|\geq 1$ the potential has
only one minimum, at $\varphi=3\pi/2$ for $k>0$ and at $\varphi=\pi/2$
for $k<0$, where the snaking orbits are captured.  The snaking
states with the lowest energy will thus have the wave vectors $k_m=\pm
R/{\ell_B^2}$.

\section{Numerical results for the infinite cylinder} \label{NRI}

\subsection{Low magnetic fields} 

In the following examples we consider a cylindrical surface with
parameters as for InAs \cite{Bringer2011}.  The effective electron 
mass and g-factor are $m_{\rm
eff}=0.023m_0$ and $g_{\rm eff}=-14.9$, respectively.  The average
electron density on the surface is $\langle n \rangle = 1.17 \times
10^{11}\ {\rm cm}^{-2}$.  We use a Rashba SOI strength $\alpha=20$
meV~nm, corresponding to a strong radial electric field which confines the
electrons on the surface, and a Dresselhaus SOI parameter $\beta=3$ meV~nm,
both inspired from flat heterostructures \cite{Winkler,Ihn}.  Indeed,
in a real sample, there is a lot of a uncertainty of these 
parameters, but in the present work we are aiming at qualitative results.
In the calculations of the spin and particle
densities we use a finite temperature $T=1$ K, sufficiently low to have
no real role, but only to avoid possible numerical artefacts produced by a pure
step-like Fermi function.

\begin{figure} 
\includegraphics [width=9 cm] {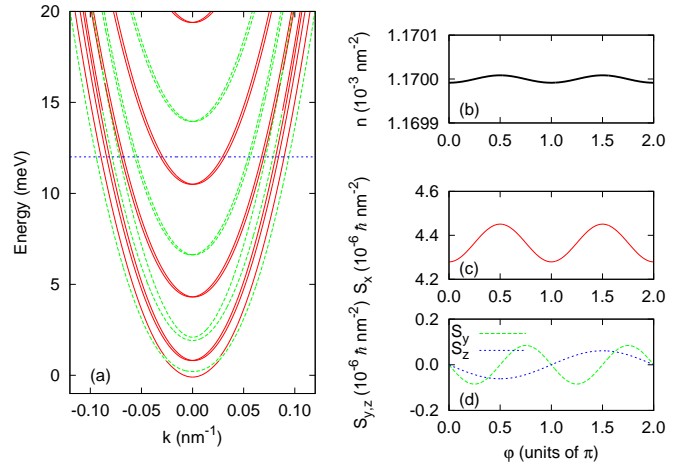}
\caption{ (a) Energy spectrum for a cylindrical surface of radius
$R=35$ nm vs. the wave vector $k$.  
The red (solid) and green (dashed) curves indicate states with a positive 
and negative spin projection along the magnetic field, respectively.  
The horizontal blue (dashed) line shows the chemical potential.  
(b) The electron density $n$, (c) the spin density
$S_x$, and (d) the spin densities $S_y$ and $S_z$ as functions of the
polar angle on the surface $\varphi$.  The magnetic field strength
$B=0.16$ T.}
\label{speden01}
\end{figure}

In Fig.\ \ref{speden01}(a) we show the energy spectrum together with the
corresponding charge and spin densities for a cylinder of radius $R=35$
nm at a low magnetic field $B=0.16$ T.  The energy spectrum
is organized in discrete energy bands (determined by the finite
circumference $2\pi R$) which are even functions of $k$, 
$E_{ak}=E_{a-k}, \ a=1,2,...$\,.  
The orientation of the spin along the direction of the magnetic field, 
positive or negative, is associated with the color used to display the 
energy dispersion curves.  
\red{Here the Zeeman energy is small, $E_Z\approx 0.86 B[\rm T]\approx 0.14$ meV, 
and hence the energy separation between the
the band pairs is determined by SOI effects.
To see this we can evaluate the energies for $B=0$ and $k=0$ using 
Eqs. (\ref{E0}) and (\ref{HR}) (neglecting $H_D$) as 
$\epsilon_{ms}=(\hbar^2/2m_{\rm eff}^2)m^2-(\alpha/R)sm$, which 
in units of meV are:
$\epsilon_{0\mp 1}=0, \ \epsilon_{\mp 1 \mp 1}=0.78, \ \epsilon_{\mp 1 \pm 1}=1.92, \ 
\epsilon_{\mp 2 \mp 1}=4.25, \ \epsilon_{\mp 2 \pm 1}=6.54$,\ etc.
These values are quite close to the energies seen in Fig.\ \ref{speden01}(a) 
at $k=0$. In the figure the degeneracies are lifted by the finite magnetic
field and the spin becomes polarized along the $x$ direction. 
The band splitting around $k=0$ is not uniform. It occurs for the lower bands, but it vanishes
at higher energies where the orbital terms proportional to $m^2$ dominate. 
} 
%

We observe a very weak variation of the electron density relative
to the mean value, around the entire circumference, Fig.\ \
\ref{speden01}(b).  This result is consistent with the classical
mechanics where according to the Bohr-van Leeuwen theorem a magnetic
field (homogeneous or not) should have no effect on the density
of electrons.  The magnetic field only translates the momenta in the
statistical distribution functions and therefore has no influence on the
equilibrium states \cite{Vleck1932}.  In our case the classical limit
corresponds to $\ell_B >> R$ or to a high electron density such that
the number of occupied subbands is much larger than one.  Therefore the
charge oscillations around the circumference have to be interpreted as
quantum effects.  The only difference between the electrons situated on
the upper and lower halves of the cylinder is the orientation of the
cyclotron motion on the surface relative to the normal direction.
Consequently the electron density has both up-down (or N-S) and
left-right (or E-W) symmetries, i.\ e.\  $n(\varphi)=n(\pi-\varphi)$ and
$n(\varphi)=n(2\pi-\varphi)\equiv n(-\varphi)$.  The weak oscillations
have thus two identical minima and two identical maxima.  In the present
example the minima occur at the $\varphi=0$ (N) and $\varphi=\pi$ (S)
regions and the maxima at $\varphi=\pi/2$ (E) and $\varphi=3\pi/2$ (W)
sides, but they may interchange for another position of the chemical
potential over the energy spectrum.

The calculated spin density corresponding to the spin direction
along the magnetic field, i.\ e.\  $S_x(\varphi)$, follows the weak
charge oscillations, with maxima and minima at the same locations,
but 
with a larger relative amplitude.  In this case 
\red{the spin density along the magnetic field} is proportional to the 
electron density, such as in a (quasi) homogeneous system. 
\red{In the SOI terms of the Hamiltonian, Eqs.\ (\ref{HR}) and (\ref{HD}),
we used the spin in cylindrical coordinates. Because of the transversal
magnetic field, in the $x$ direction, we prefer to show in Fig.\ \ref{speden01}
all spin densities in cartesian coordinates.} 
The spin density perpendicular to the
magnetic field, in the $y$ direction, i.\ e.\ transverse to the net
motion, is determined by the Rashba SOI, and it is an odd function
of angle, $S_y(\varphi)=-S_y(-\varphi)$, and also antisymmetric on
the lateral sides of the cylinder w.r.t $\varphi=\pm \pi/2$,
$S_y(\varphi)=-S_y(\pi-\varphi)$. The effect of the Dresselhaus SOI is
to tilt the spin in the $z$ direction, i.\ e.\ along the net motion,
but with the symmetries $S_z(\varphi)=-S_z(-\varphi)=S_z(\pi-\varphi)$.
\red{ 
In the example shown in Fig.\ \ref{speden01} the oscillations of the
spin densities in all spatial directions $(x,y,z)$ are comparable in
amplitude, but the average values of $S_y(\varphi)$ and $S_z(\varphi)$
vanish. This is due to the circular symmetry of the SOI electric field,
as opposed to the fixed direction of the Zeeman field.  }

\subsection{High magnetic fields} 

\begin{figure} 
\includegraphics [width=9 cm] {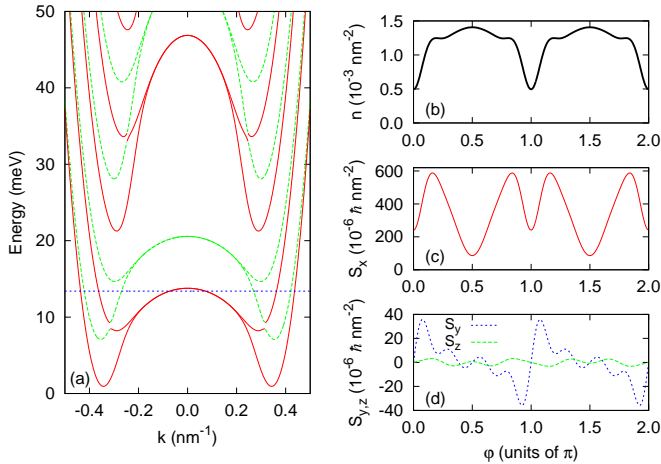}
\caption{ (a) Energy spectrum for a cylindrical surface of radius
$R=35$ nm vs. the wave vector $k$.  
The red (solid) and green (dashed) curves indicate states with a positive 
and negative spin projection along the magnetic field, respectively.  
The horizontal blue (dashed) line shows the chemical potential.  
Right: (b) The electron density $n$,
(c) the spin density $S_x$, and (d) the spin densities $S_y$ and $S_z$ as functions of
the polar angle on the surface $\varphi$.  The magnetic field strength
$B=7$ T.}
\label{speden03}
\end{figure}

In the next example we increase the magnetic field to $B=7$ T and show
the results in Fig.\ \ref{speden03}.  In this case the magnetic length
$\ell_B=9.7 \ {\rm nm} << R=35 \ {\rm nm}$ and therefore the electrons
situated at the N or S regions have almost localized wave functions
around $\varphi=0$ and $\varphi=\pi$, respectively, where they experience a
nearly constant magnetic field.  Such states are in fact Landau levels
which can be seen in the center of the energy bands, around $k=0$
\cite{Ferrari2008,Bellucci2010}.
In Fig.\ \ref{speden03}(a) the chemical potential is slightly below the
lowest Landau level.  Each Landau level is actually split in
two sublevels because of the two equivalent regions N and S. For such
a large magnetic field the two sublevels almost merge 
and hence the Landau level becomes almost degenerate.  The Zeeman energy is now
6 meV and the next band top above the chemical potential coincides with
the same ground Landau level, but with the opposite spin projection in the
$x$ direction.  The states which
are not around the center of the energy bands have wave functions on the
lateral sides of the cylinder, possibly in the snaking areas.  The snaking
states with the lowest energy correspond to the lowest side minima of
the lowest band and have the wave vectors which minimize the effective
potential (\ref{Vmag}), $k_m=\pm R/{\ell_B^2}=\pm0.37$  {\rm nm}$^{-1}$.

\red{An interesting feature of the energy spectrum is the non
monotonous band dispersion.  The energy bands have local minima and maxima
which are expected to produce conductance steps. Each time the Fermi energy
crosses a minimum the coductance increases by one unit, and each time
it crosses a maximum, it decreases by one unit of $e^2/h$.  As a result
``anomalous'' conductance steps should occur,  
taking as a reference the simple staircase corresponding to displaced
parabolas around $k=0$ \cite{Moroz2000}. }


We notice another interesting feature of the energy spectrum.  The Landau
states increase in energy proportionally with the strength of the magnetic
field, whereas the minima of the magnetic potential are always zero.
This means that with increasing magnetic field the tops of the
bands at $k=0$ rise whereas the side minima are relatively stable.
Consequently the electron density develops minima at the N and S poles and
maxima at the E and W sides, as can be seen in Fig.\ \ref{speden03}(b).
Concerning the spin density, because the Zeeman gap is now large, the $x$
projection does not follow the electron density like in the low field
limit, Fig.\ \ref{speden03}(c).

\begin{figure} 
\includegraphics [width=9 cm] {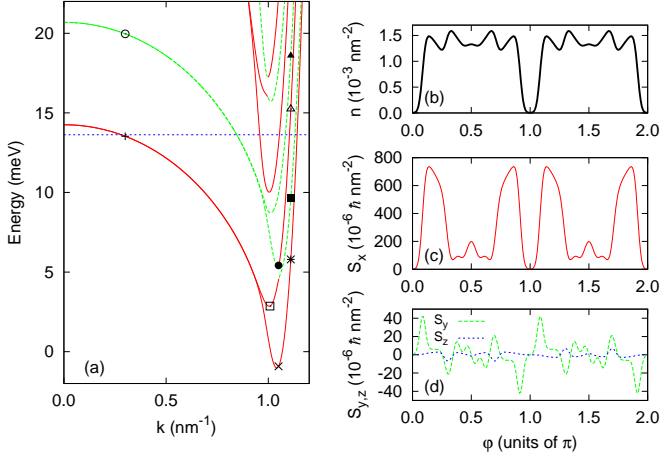}
\caption{ (a) Energy spectrum for a cylindrical surface of radius $R=100$
nm vs. the wave vector $k$.  In this case only the energies with $k>0$
are shown.  Selected energies are marked with various point symbols to
be correlated with Fig.\ \ref{probden06}.  
The red (solid) and green (dashed) curves indicate states with a positive 
and negative spin projection along the magnetic field, respectively.  
The horizontal blue (dashed) line shows the chemical potential.  
(b) The electron density $n$,
(c) the spin density $S_x$, and (d) the spin densities $S_y$ and $S_z$ as functions
of the polar angle on the surface $\varphi$.  The magnetic field strength
$B=7$ T. }
\label{speden06}
\end{figure}

Next we show in Fig.\ \ref{speden06} results for the same magnetic field
strength as before, $B=7$ T, but for a cylinder with radius $R=100$ nm.
In this case the Landau states extend over wider surfaces on the
top and bottom of the cylinders, or in other words on a larger interval of
center coordinates $k\ell_B^2$.  However, with the chemical potential
below the lowest Landau state, the electron density totally vanishes
at the N and S poles.  Increasing the magnetic field, 
or reducing the electron density, the depleted regions will expand, and 
eventually the electrons will concentrate only along thin lateral channels.

The spin-$x$ density has pronounced peaks matching the lateral maxima
of the charge distribution which correspond to electrons in the lowest
pair of (almost degenerate) energy bands, and thus all with spin ``up''.
The secondary maxima at $\varphi=\pi/2$ and $\varphi=3\pi/2$ are given by
the snaking electrons.  But due to the heavy bending of the energy bands,
they now reside in the lowest three pairs of energy bands, the first
two (more or less) with compensated spin, but the third again with spin
``up''.
%
\begin{figure} 
\includegraphics [width=8 cm] {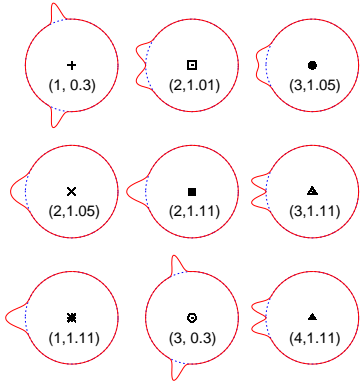}
\caption{The probability density for selected states is
drawn with the continuous red line. The dashed blue line indicates the
contour of the cylinder.  The selected states are marked
with the same point symbols in the energy spectrum and in the center of
each circle, respectively.  The quantum numbers $(a,k)$, i.\ e.\ the band
index and the wave vector, are also shown.  The states with maxima at
N and S behave like cyclotron orbits.  The states concentrated at W are
snaking orbits.  In principle all states with a sufficiently large $k$
are snaking with a number of peaks related to the energy quantization
in the effective potential well created around the W point.}
\label{probden06}
\end{figure}

The probability density around the circular contour, for several
representative single particle states marked with point symbols, is
shown further in Fig.\ \ref{probden06}.  For example, the state marked
with {\small $+$} belongs to the band $E_{1k}$, has energy 13.5 meV
and wave vector $k=0.3 \ {\rm nm}^{-1}$, and it is of cyclotronic type.
The probability density has two symmetric peaks slightly tilt towards
the W side as shown in the top-left example of Fig.\ \ref{probden06}.
The state $E_{2k}$ is next on the energy scale, but practically
with the same energy as $E_{1k}$, and with the same probability density
(not shown).  Going along the lowest band we show the  state marked
with {\small $\times$}, which on our $k$-grid is the nearest to the
band minimum $k_m=R/{\ell_B^2}=1.06 \ {\rm nm}^{-1}$.  Here the two
probability-density peaks of the former state merge and the state
becomes of a snaking type.  Increasing further $k$ within the same
band we find only snaking states with one distribution peak, like
for example the state marked with {\small $+\hspace{-2.5mm}\times$}.
In the second energy band we show the state {\small $\boxdot$} which is
in transition from cyclotronic to snaking type, and the state {\small
$\blacksquare$} which is again snaking.  In this region of the spectrum
this later state differs from the state {\small $+\hspace{-2.5mm}\times$}
only by the spin orientation.  In the absence of the SOI a band crossing
would occur close to the state {\small $\CIRCLE$}. Here, i.\ e.\ on the
third energy band, the snaking states have a double peak distribution
which persists for all higher $k$'s, e.\ g.\ for the state \ {\small
$\cdot\hspace{-1.90mm}\triangle$}. This double peak shows that this 
snaking state corresponds to a bound state in the effective magnetic 
potential which is no longer the ground state, but the first excited 
state, i.\ e.\ with one extra oscillation. The same type of state is 
the one marked with {\small $\blacktriangle$}, though it is different 
from the former by the spin state. A detailed semiclassical description 
of such states is given in Ref. \cite{Bellucci2010}.

\subsection{Evanescent states} 

\begin{figure} [H]
\begin{center}
\includegraphics [width=9.5 cm] {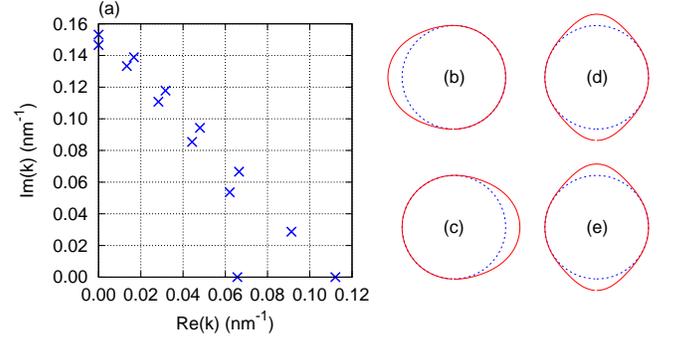}
\end{center}
\vspace{-10mm}
\caption{Evanescent states. Left: (a) Complex wave numbers for a cylindrical 
surface of radius $R=35$ nm in a magnetic field $B=2$ T for an energy $E=1.5$ meV. 
Right:  Angular distribution for selected wave numbers: 
(b) $k=(0.091, \pm 0.029)$, 
(c) $k=(-0.091, \pm 0.029)$, 
(d) $k=(0, \pm 0.147)$, 
(e) $k=(0, \pm 0.153)$. 
}
\label{figx}
\end{figure}

To end this section, we analyze the behavior of the eigenstates
when the wave number is allowed to take complex values. These are
the so-called evanescent states and they \red{become} important close to
interfaces or inhomogeneities in the cylinder. \red{These are virtual states 
which participate in transport.} 
Fig.\ \ref{figx} displays
the distribution of evanescent wave numbers in the complex plane for a
particular set of parameters.  They have been obtained numerically using
the method of Ref. \cite{Serra2007}.  The figure only displays the first
quadrant of the complex plane since the wave numbers are distributed
symmetrically around the real and imaginary axis. That is, if $k$ is an
allowed complex wave number, then $\pm k$ and $\pm k^*$ are also allowed.
For simplicity the spin is not included in this analysis. 
\vspace{-10mm}
As with the purely propagating states discussed above, Fig.\ \ref{figx}
shows that for relatively small ${\rm Im}(k)$ the snaking character is along
W for ${\rm Re}(k)>0$  and along E for ${\rm Re}(k)<0$, independently on the sign
of ${\rm Im}(k)$.  However, as ${\rm Im}(k)$ increases there is a qualitative change
and the modes tend to be localized around N and S, \red{like cyclotron orbits}. As a limiting case, for
the purely imaginary $k$'s there is an exact localization around N and S.
The ratio $R/\ell_B \approx 1.9$ here, and hence the localization range is wider 
than in Fig.\ \ref{speden06}(b) where $R/\ell_B\approx 10$.

\section{Cylinder of finite length} \label{NRF}

\subsection{Depletion of the electron density} 

As a second model of a cylinder, perhaps more realistic for the experimental
research, we consider one with a finite length $L$.  We impose hard wall boundaries 
in the $z$ direction at  $z=0$ and $z=L$ and hence the 
previous axial plane waves $\langle z|k\rangle$ must be replaced by the discrete set
\begin{equation}
\langle z |p\rangle=\sqrt{\frac{2}{L}} \sin \frac{p\pi z}{L}\ , \ p=1,2,...\ .
\end{equation}
In this case the matrix elements of the total Hamiltonian
couple together all degrees of freedom, i.\ e.\ the angular, the
axial, and the spin motions.  However, for regions far from the edge
we expect results comparable with the infinite cylinder if $R << L$.
An effective potential for the snaking states cannot be easily defined,
but the presence of such states can still be identified.

The matrix elements of the Hamiltonian with boundary conditions are shown
in the Appendix A.  We performed numerical calculations for a cylinder of
length $L=336$ nm and a ten times smaller radius $R=33.6$ nm.  The number
of electrons on the surface is $N=10$ or $N=90$, as further specified.
$N=90$ corresponds to a surface density of $1.27 \times 10^{11} \ {\rm
cm}^{-2}$, which is quite realistic.  In the experimental core-shell
samples the number or electrons can be adjusted with gates attached to the
surface of the cylinder \cite{Blomers2011}.  The Rashba and Dresselhaus
SOI strengths are the same as for the infinite cylinder, $\alpha =
20$ and $\beta = 3$ meV~nm, respectively, unless otherwise specified.
The numerical results were convergent with a basis set consisting of
up to 2738 states.  Due to computational limitations we used magnetic
fields of lower values than for the infinite cylinder only up to 4 T.

\begin{figure} 
\begin{center}
\includegraphics[width=7cm]{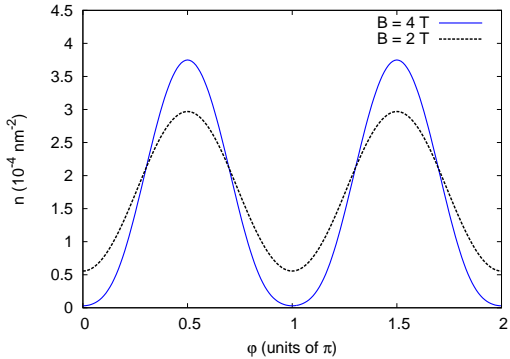}
\end{center}
\caption{ Electron density calculated close to the 
middle of the cylinder, at $z=160$ nm from one edge.  
With $N=10$ electrons the Fermi energy decreases and so the N and S
regions may become totally unpopulated for a magnetic field strength like
$B=4$ T.  
}
\label{finite1}
\end{figure}

We first show in Fig.\ \ref{finite1} the density of electrons around the
circumference, in a point situated somewhere in the middle of the length
of the cylinder, at $z=160$ nm.  In order to obtain conditions for the
depletion of the N and S regions at magnetic fields reachable by our
numerical approach we chose the number of electrons $N=10$.  As expected
from the previous calculations for the infinite length, since $\l_B \ll
L$, the density profile for $B=2$ T shows pronounced dips, reaching zero
at $B=4$ T, at the N and S poles ($\varphi=0, \pi, 2\pi$), whereas the
peaks show the concentration of electrons around the snaking orbits,
at E and W sides ($\varphi=\pi/2, 3\pi/2$).

\subsection{Torsion of the electron density} 

\begin{figure} 
\begin{center}
\includegraphics[width=4.35cm]{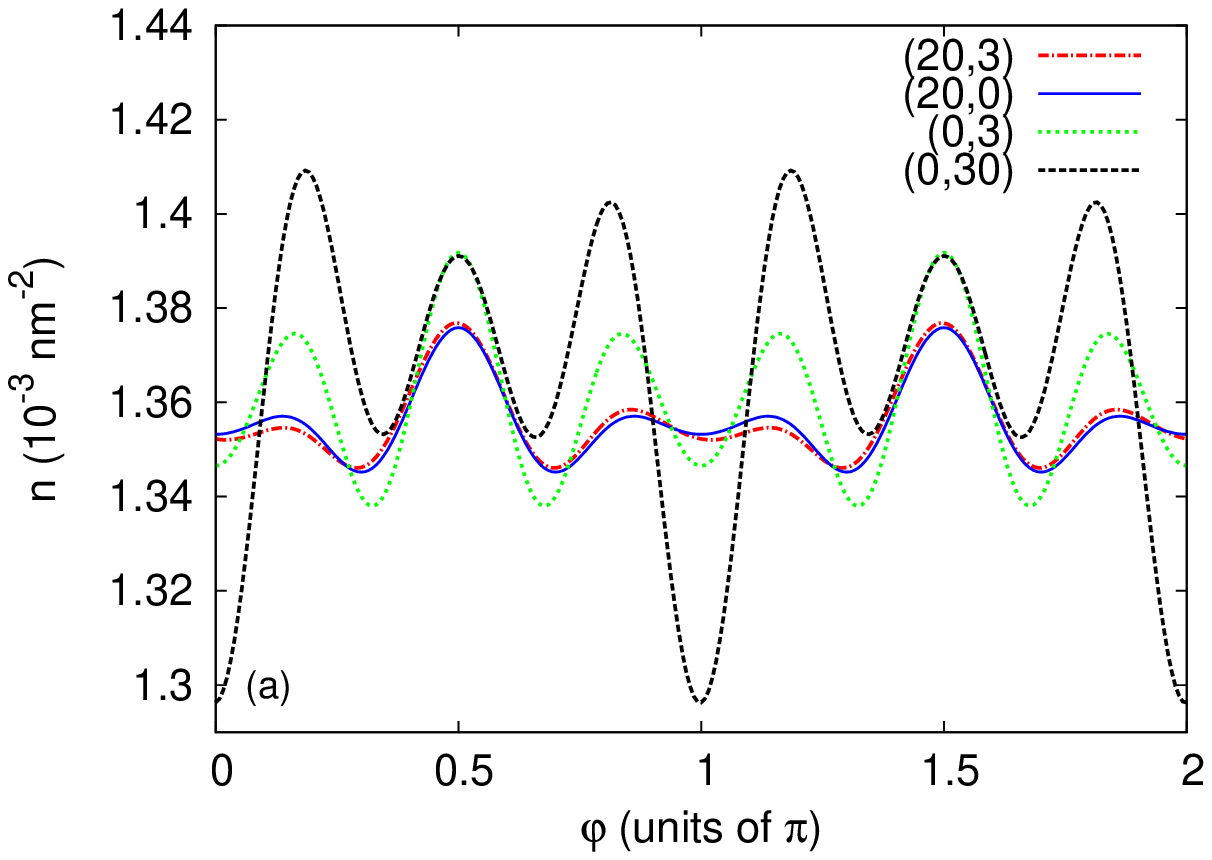}
\includegraphics[width=4.35cm]{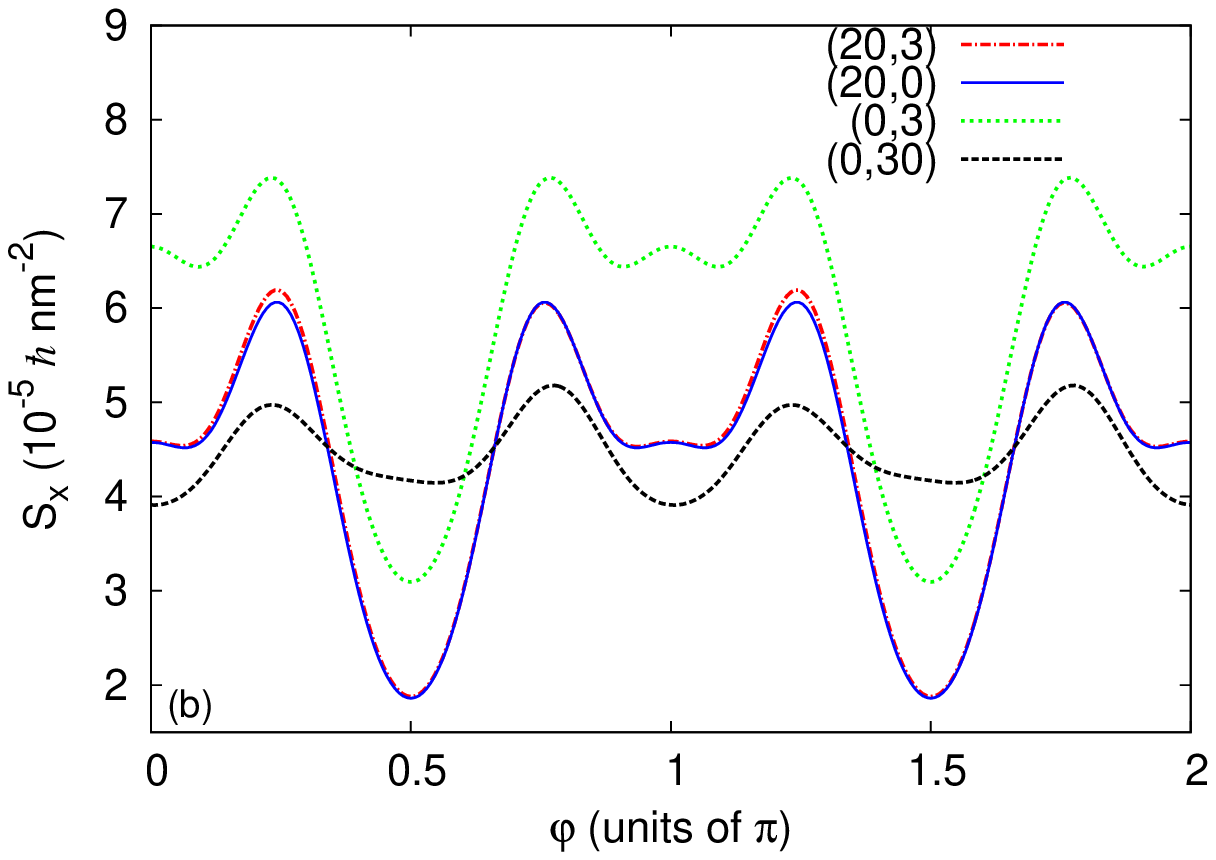}
\includegraphics[width=4.35cm]{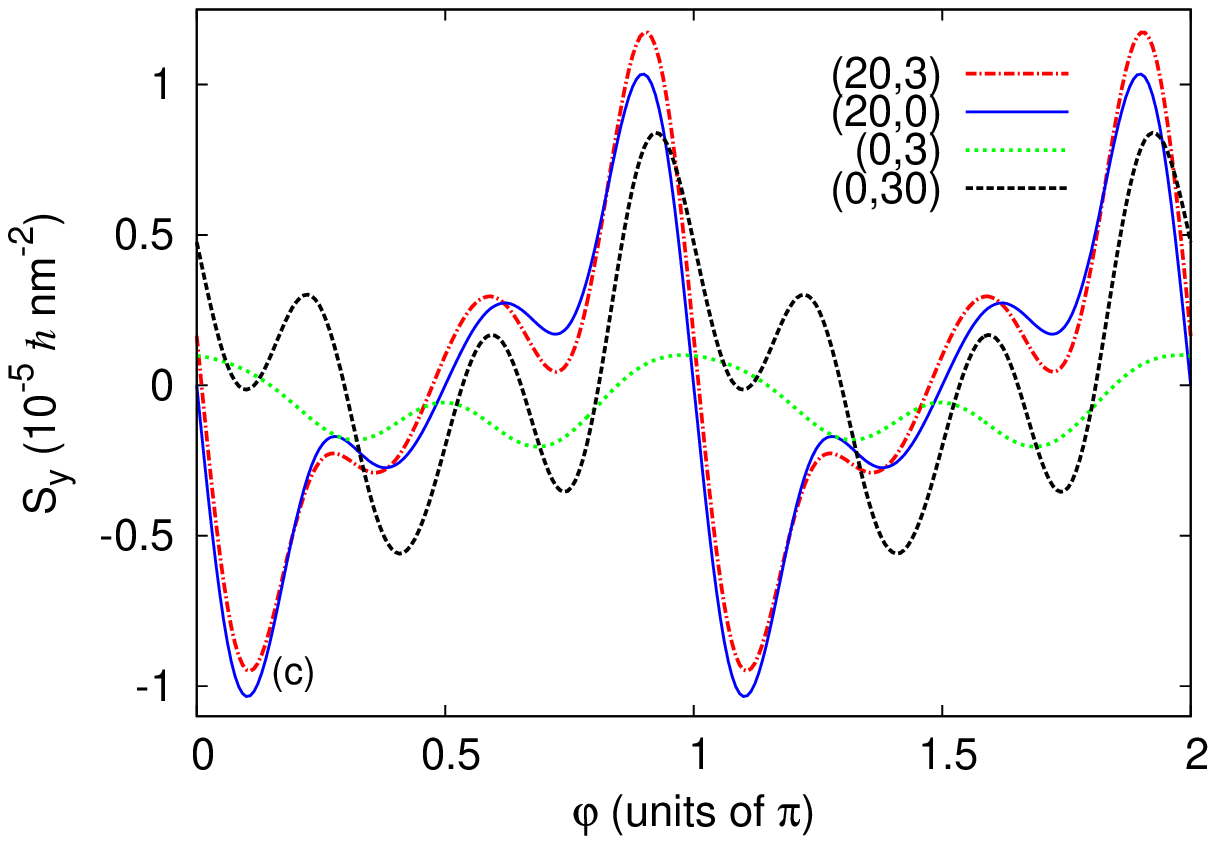}
\includegraphics[width=4.35cm]{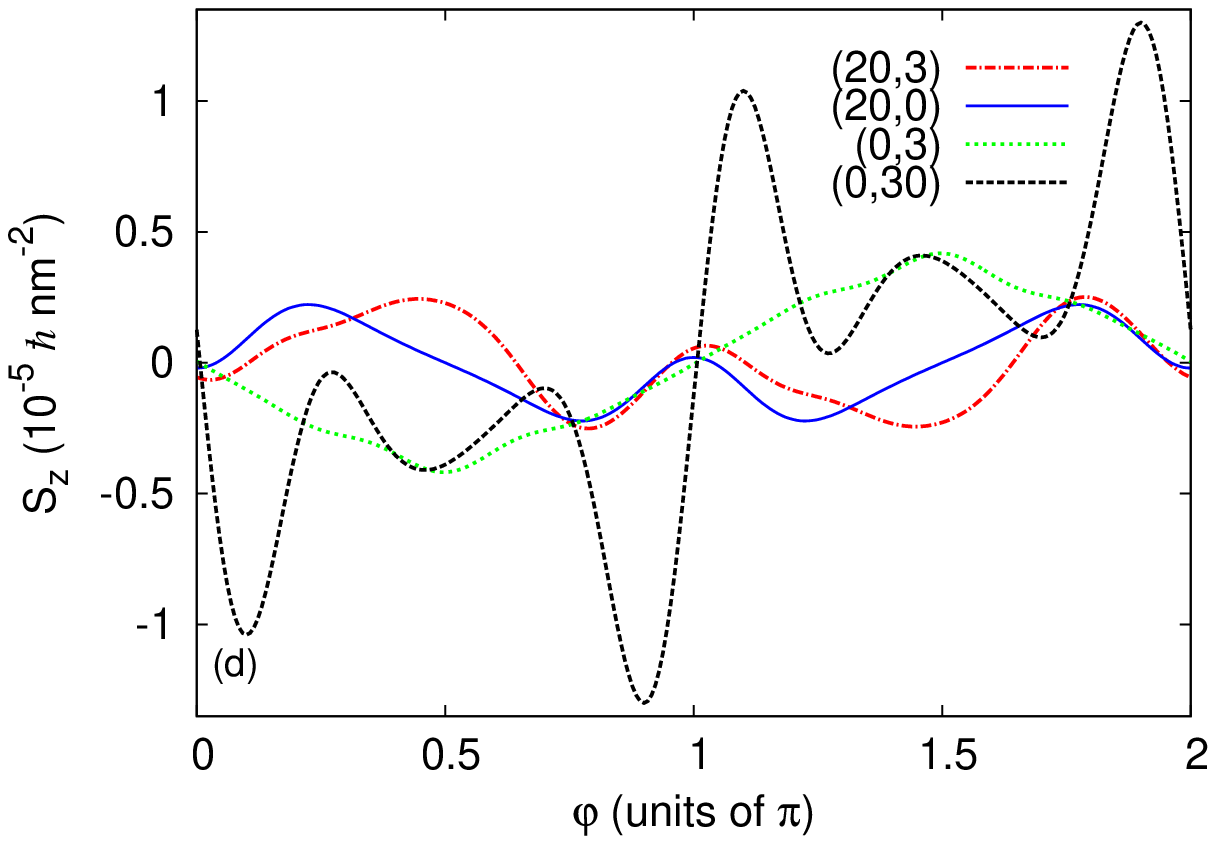}
\end{center}
\caption{Charge and spin densities at $z=160$ nm from the edge of a 336 nm
long cylinder, for $B=2$ T and $N=90$ particles. Each curve corresponds
to a combination of SOI parameters $(\alpha,\beta)$ indicated in the
legend (in units of meV nm).  (a) The electron density is slightly
asymmetric relatively to $\varphi=0$ and $\varphi=\pi$ for
$(\alpha,\beta)=(20,3)$ because of the Dresselhaus SOI.  It becomes
symmetric for $\beta=0$.  The asymmetry is not visible in the graph
for a small $\beta$ like 3 meV nm, but it becomes clear for a larger $\beta$
like 30 meV nm. (b, c, d) The analog results for the spin densities
$S_{x,y,z}$ respectively.  The asymmetry is more clearly seen in $S_{y}$ and 
$S_{z}$.} 
\label{finite2}
\begin{center}
\vspace{-10mm}\hspace{-5mm}\includegraphics[width=10cm]{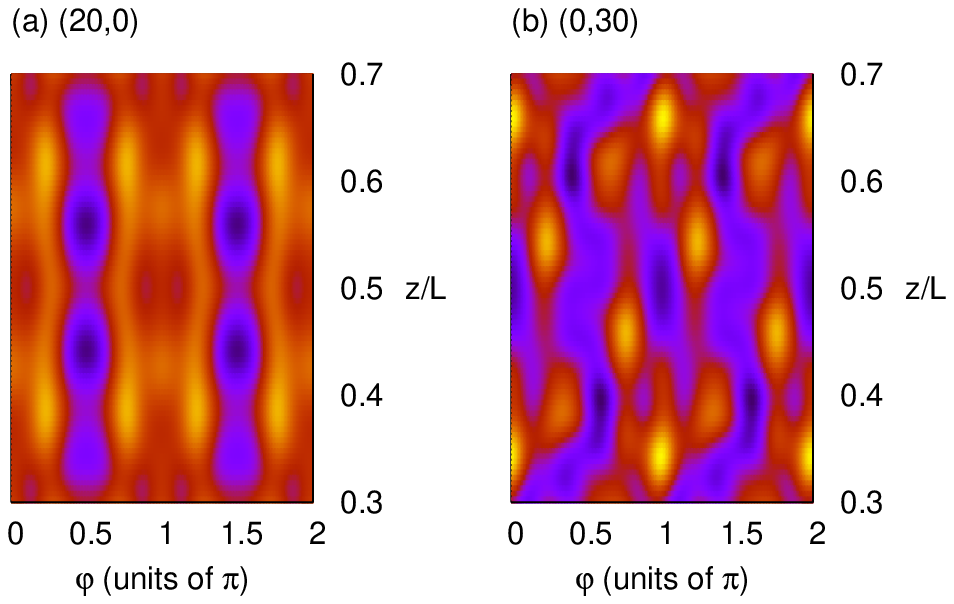}
\end{center}
\vspace{-10mm}\caption{Qualitative plots of the spin densities $S_x$ 
for a cylinder of length $L=336$ nm, with $N=90$ electrons and $B=2$ T, on a length 
interval of $0.4L$ around the center. 
(a) $\alpha=20$ meV nm, $\beta=0$ and (b) $\alpha=0$ and $\beta=30$ meV nm.
Bright colors correspond to maxima and dark to minima.}
\label{torsion}
\end{figure}

Next, in Fig.\ \ref{finite2} we show the charge and spin densities at
the same location on the cylinder as before, $z=160$ nm, but for an
increased number of electrons, $N=90$, i.\ e.\ including now states with
higher energy.  Dips at N and S and peaks at E and W poles can still
be observed in the electron density, although separated by additional
fluctuations, Fig. \ref{finite2}(a).  But particularly interesting now
is the left-right asymmetry of the electron density relative to the N-S
(or $x$) axis, around $\varphi=\pi$ and $\varphi=0$. For the SOI
parameters used until now $(\alpha, \beta)=(20,3)$ meV nm, the effect is
weak (at $B=2$ T).  The reason for this asymmetry is the Dresselhaus
SOI.  To demonstrate that we also show the electron density with Rashba
SOI only, $(\alpha, \beta)=(20,0)$ meV nm, which is symmetric. 
For Dresselhaus SOI only the asymmetry is very small, unobservable
in Fig. \ref{finite2}(a) for $\beta=3$ meV nm, but it becomes visible
for a larger value like $\beta=30$ meV nm.

The Dresselhaus SOI did not break the symmetry of the electron
distribution on the infinite cylinder. For the cylinder with a finite
length, at any $z$ location,  we obtain that symmetry only for $\beta=0$,
i.\ e.\ the N-S reflection $n(\varphi,z)=n(\pi-\varphi,z)$, and the E-W
reflection as well, $n(\varphi,z)=n(-\varphi,z)$, and consequently the
inversion in any cross section, i.\ e.\  $n(\varphi,z)=n(\varphi+\pi,z)$.
In addition, we also have reflection symmetry along the cylinder,
$n(\varphi,z)=n(\varphi,L-z)$, for any angle $\varphi$.  If $\beta\neq 0$ 
we now obtain a reduced symmetry: the inversion in any cross section
survives, whereas the N-S and E-W reflections must be combined with
longitudinal reflection, i.\ e.\  $n(\varphi,z)=n(\pi-\varphi,L-z)$, and
$n(\varphi,z)=n(-\varphi,L-z)$, respectively.  In Fig.\ \ref{finite2}(a)
we show density profiles at $z=160$ nm.  The profiles at the symmetric
point w.r.t. the center of the cylinder, $L-z=176$ nm, correspond
to a reflection relatively to the $\varphi=\pi$ axis.  Only exactly in the middle
of the cylinder the N-S and E-W reflection symmetries are restored and the
density profile becomes of the same type as for the infinite cylinder.
Hence, the Dresselhaus SOI creates a torsion of the electron density on the
finite cylinder, with opposite twist angles relatively to the center. We
have checked numerically that when increasing the length of the cylinder
the torsion weakens on a certain finite $z$ interval around the center,
and so the density profile slowly evolves towards that for the infinite cylinder.

In the presence of both types of SOI the torsion of the charge density
appears amplified.  A more or less related effect has been discussed
in the recent literature for quantum rings in a magnetic field
perpendicular to the plane of the ring.  In that case the combination
of the two types of SOI breaks the rotational symmetry in the spin
space and leads to a self-consistent deformation of the charge and spin
densities \cite{Sheng2006,Nowak2009,Daday2011}.  But in that case both
Dresselhaus and Rashba SOI must be present, whereas for the deformation
presented here only the Dresselhaus SOI is sufficient.  

In Fig.\ \ref{finite2}(b,c,d) we show the spin densities corresponding to
the spin projections $S_{x,y,z}$.  The features of the spin density $S_x$
are similar to those of the particle density, i.\ e.\ the symmetries
and the torsion.  They are also reproduced by the spin densities
$S_{y,z}$, possibly with a sign change of the density, like for the
infinite cylinder.  But unlike in that case, where Rashba SOI may tilt
the spin only in the $y$ direction and Dresselhaus SOI only in the $z$
direction, on the finite cylinder each one can tilt it in both $x$
and $y$. The reason is that on the finite cylinder the electrons have
a combined circular and longitudinal net motion.  Both $S_y$ and $S_z$
spin projections are generally more sensitive to the torsion created by
the Dresselhaus SOI as can be seen in Fig.\ \ref{finite2}(c,d) even for
$\beta=3$ meV nm.  To illustrate better the effect we display in Fig.\
\ref{torsion} a color map of the spin density $x$, which does not change
direction (for the parameters used) and is therefore simpler.  We will
return to the torsion aspect in the next subsection in the context of
the current density.

\subsection{Current distribution} 

\begin{figure} 
\begin{center}
\includegraphics[width=7cm]{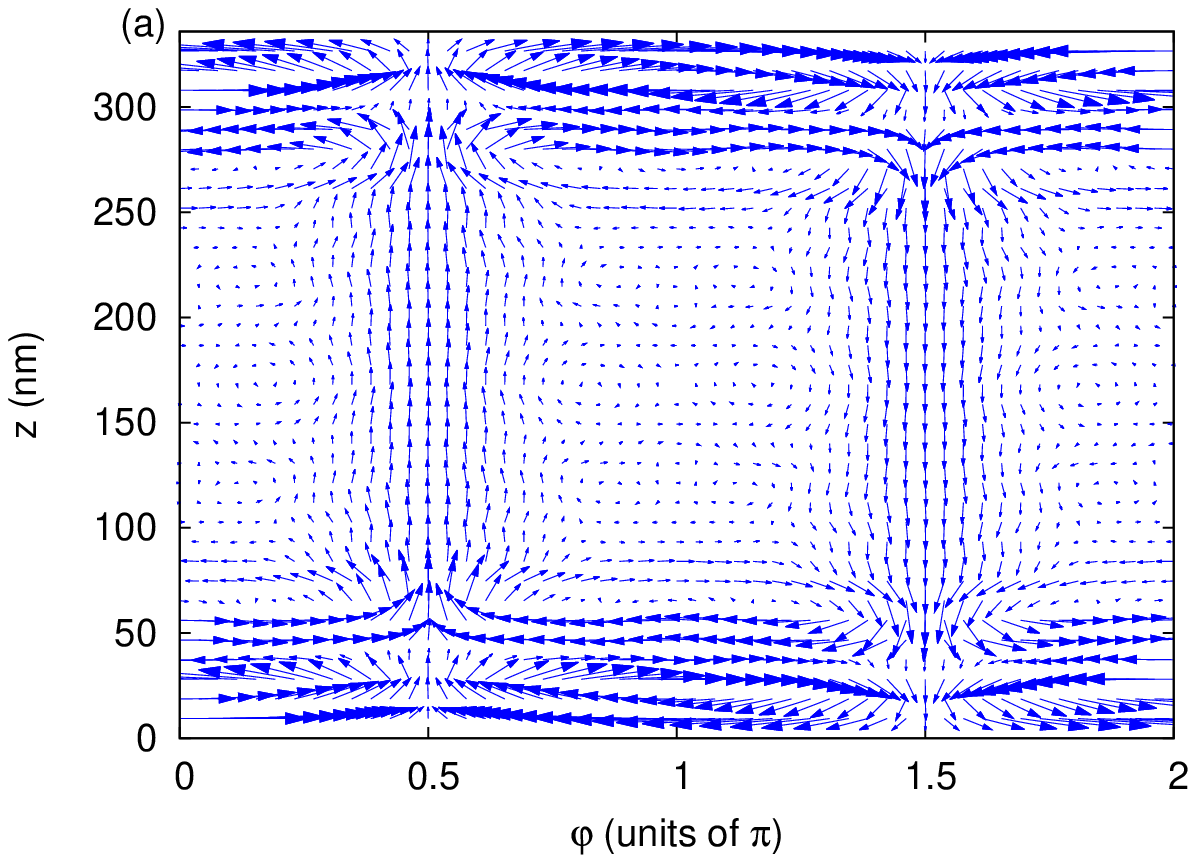}
\includegraphics[width=7cm]{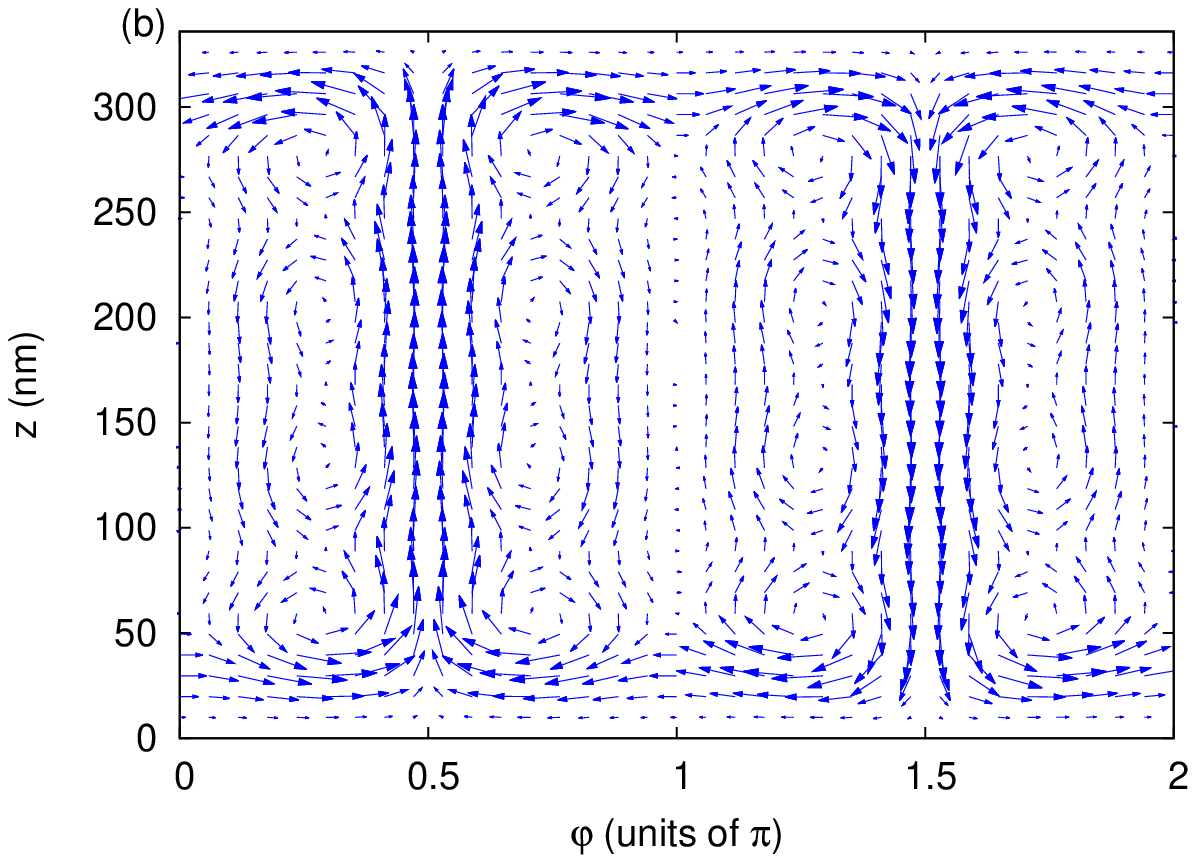}
\includegraphics[width=7cm]{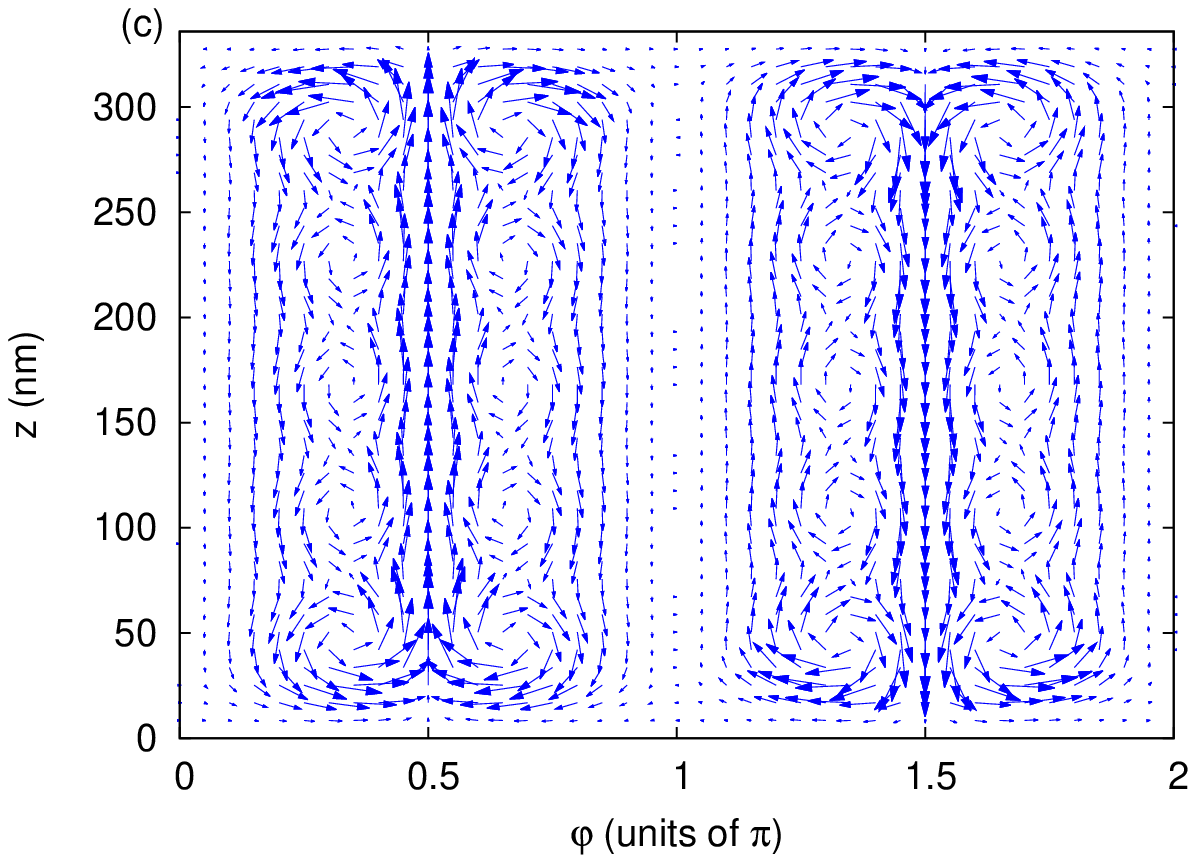}
\end{center} \caption{Equilibrium current density on the surface of
the cylinder.  In all cases $(\alpha,\beta)=(20,3)$ meV nm.  (a) $B=2$
T and $N=90$ electrons. (b) $B=2$ T and $N=10$. (c) $B=4$ T and $N=10$.}
\label{finite3}
\end{figure}

Finally, in Fig.\ \ref{finite3} we illustrate the equilibrium current
density on the surface of the cylinder at high magnetic fields.  We 
see the coexistence of cyclotron and snaking trajectories.  For a low
electron number like $N=10$ the depletion of the N and S regions is clear,
to be compared with Fig.\ \ref{finite1}.  Nearly straight propagation
occurs at the E and W sides.  It is also interesting to notice the
vortices occurring at the edges and how the edge currents are absorbed
or ejected by the snaking states.  In a way the snaking orbits may also
be seen as very large loops of current surrounding the whole cylinder,
which would correspond to large, special cyclotron orbits associated with
a very low net magnetic field experienced on the average by the electrons
trapped along the zero-radial-field lines.  One observes in Fig.\
\ref{finite3}(a,b) how the snaking currents split at the upper edge at
$\varphi=\pi/2$ and join back at $\varphi=3\pi/2$, and vice versa at the
lower edge.  When the depletion of the N and S regions occurs, like in
Fig.\ \ref{finite3}(c), the edge currents connecting the snaking states
tend to vanish and therefore the E and W halves of the cylinder tend to
become separated or independent.

The twisted symmetry is also present in the current and spin
distributions, but not clearly visible in these figures.  We have checked
with the numerical calculations that, in order to suppress the torsion,
it is sufficient to remove the first term of Eq.\ (\ref{HD}), proportional
to $\sigma_{\varphi}p_{\varphi}+p_{\varphi}\sigma_{\varphi}$.  This term
describes the SOI at the edges of the cylinder associated with the
currents connecting the snaking states over the N and S regions.  Those
currents are basically given by the tangential momentum $p_{\varphi}$.
Therefore, in the depletion regime, when the edge currents vanish, the
torsion disappears.  We did confirm this fact with several numerical
tests.  The fact that $\sigma_{\varphi+\pi}=-\sigma_{\varphi}$ indicates
that the spin currents should have the same twisted symmetry as the charge
and spin densities.  A more detailed study of the spin texture and spin
currents is beyond the scope of this work; it needs to be done separately
\red{and compared to the results known for flat quantum wires \cite{Bellucci2003}.}

We notice the different behavior of the two types of SOI at the spatial
inversion $(x,y,z)\to (-x,-y,-z)$, or $(\varphi,z)\to (\varphi+\pi,L-z)$,
which leaves invariant $H_R$, but changes the sign of $H_D$.  Hence,
with $\beta\to -\beta$ or $B \to -B$  we obtain all the distributions
flipped along the $z$ direction.  This is essentially the reason for
the torsion effect on the cylinder of finite length, and the suppression on 
the infinite cylinder where the inversion of the $z$ axis has no real meaning.


\section{Conclusions} \label{CON}

We calculated electronic quantum states on the surface of a cylinder of
radius of 30-100 nm having in mind a core-shell nanowire.  We used a
uniform magnetic field transverse to the cylinder and we analyzed 
the charge, spin, and current densities.  For a strong magnetic field,
like a few tesla, when the magnetic length is smaller than the radius
of the cylinder, the energy spectrum can be considered a combination of
locally closed cyclotron orbits and long, open, snaking states.
To our knowledge other studies of such a system focused mostly on the
character of the individual electronic states, cyclotronic or snaking,
but not on the net effect of these states in a many-electron system.

We showed that if the Fermi energy is sufficiently
low, corresponding to realistic electron densities of the order of
$10^{11}\ {\rm cm}^{-2}$, the distribution of the electrons around the
circumference of the cylinder may totally vanish at the top and bottom
regions and the electrons will concentrate on the sides of the surface.
This highly anisotropic distribution may have interesting effects on the
magnetoresistance and possibly on other transport or optical properties
of core-shell nanowires whenever the local electron density or the local
current density plays a role.  For example if such a wire can function
as a nano antenna then the directivity may be easily controlled by an
external magnetic field.  We also obtained the peculiar equilibrium 
current distribution on the surface of the finite cylinder which 
is relevant for the magnetization properties of such systems.

We also calculated the anisotropic spin distribution around the cylinder
at these high fields in all spatial directions, using standard models for
the Rashba and Dresselhaus spin-orbit interactions.  In particular, for a
cylinder of finite length, the Dresselhaus SOI breaks the left-right and
up-down symmetry of the charge and spin distributions, adding to them an
inversion along the length. This results in a torsion of the distributions
with opposite twist angles relatively to the center of the cylinder.
The effect is produced by the Dresselhaus SOI, through the azimuthal currents, 
and it may be amplified by the Rashba SOI.  The effect may be observable
in transport or in magnetization measurements.


\acknowledgements
This work was supported by the Research Fund of the University of Iceland, 
the Icelandic Research Fund, and by grant FIS2011-23526 (MINECO).
Stimulating discussions with Thomas Sch\"apers, Mihai Lepsa, 
and Klaus-J\"urgen Friedland are very much appreciated.

\appendix*

\section{The matrix elements of the Hamiltonian}

In this appendix we show the matrix elements of the Hamiltonian (\ref{H}).
The orbital component is denoted as $H_O=H_0+H_B$.  We use the characteristic 
rotational energy $E_0=\hbar^2/2m_{\rm eff}R^2$, the cyclotron energy 
$\omega_c=eB/m_{\rm eff}$, the magnetic length $\ell_B=\sqrt{\hbar/eB}$. 

For the infinite cylinder the basis is Eq.\ (\ref{basis}), all matrix 
elements are diagonal with respect to the wave vector $k$, and we obtain:

\begin{eqnarray}
&& \left<mk s \left|H_O \right|m'ks' \right> = 
\bigg\{
\frac{E_0}{2}\left[m^2 + (k R)^2\right]\delta_{mm'} - \nonumber \\
&& \frac{i\hbar\omega_c kR}{2} \delta_{m,m'+1} + 
\frac{\hbar\omega_c R^2}{8\ell_B^2}\left(\delta_{mm'} - \delta_{m,m'+2}\right) 
\bigg\}
\delta_{ss'} \nonumber  + {\rm H.\ c.}, \nonumber
\end{eqnarray}
where H.\ c. stands for the Hermitean conjugation.
The matrix elements of the Zeeman Hamiltonian (\ref{HZ}) are
\begin{equation}
 \left<m'ks' \left|H_Z \right|mks \right> = -\frac{1}{2}g_{\rm eff}\mu_B B\delta_{mm'}
\delta_{-s,s'}.
\nonumber
\end{equation}
The SOI terms (\ref{HR}) and (\ref{HD}) have the following matrix elements:
\begin{eqnarray}
&& \left<mks \left|H_R \right|m'ks' \right> = 
-\frac{\alpha}{2R}ms\delta_{mm'}\delta_{ss'}  + \nonumber \\
&& \frac{i\alpha k}{2}(1-s)\delta_{m,m'+1}\delta_{-s,s'} + \nonumber \\
&& \frac{\alpha R}{4\ell_B^2} \left[ (1-s)\delta_{m,m'+2} - \delta_{mm'} \right] \delta_{-s,s'}  
    + {\rm H.\ c.}, \nonumber \\ 
\nonumber \\
\nonumber \\
&& \left<mks \left|H_D \right|m'ks' \right> = 
    -\frac{\beta k}{2}s\delta_{mm'}\delta_{ss'} + \nonumber \\
&&   \frac{i\beta}{4R} (1-s)(2m-1) )\delta_{m,m'+1} \delta_{-s,s'} + \nonumber\\
&&    \frac{i\beta R}{2\ell_B^2} \delta_{m,m'+1} s\delta_{ss'}
    + {\rm H.\ c.} \nonumber
\end{eqnarray}

For the finite cylinder, with hard-wall boundary conditions in the $z$
direction, the basis is $|mps\rangle$ with integer $p$.  In this case we obtain:
\begin{equation}
 \left<mps \left|H_O \right|m'p's' \right> = \left\{ 
  \begin{array}{l l}
    \bigg\{\frac{E_0}{2}\left[m^2 + \frac{(p\pi R)^2}{L^2} \right]\delta_{mm'} +\\
    \frac{\hbar\omega_c R^2}{8\ell_B^2}\left(\delta_{mm'} - \delta_{m,m'+2}\right) \bigg\}\delta_{ss'} \\ 
     + \ {\rm H.\ c.} \hspace{5 mm} \text{if $p=p',$ }\\
    \\ \\
    -2\hbar \omega_c\frac{R}{L}\frac{pp'}{p^2 - p'^2}\delta_{m,m'+1}\delta_{ss'} \\ 
    + \ {\rm H.\ c.} \hspace{5 mm} \text{if $p+p'$ odd,}\\
   \\ \\
 0 \hspace{12 mm} \text{otherwise.} \\  
  \end{array} \right.
\nonumber
\end{equation}
The matrix elements of the Zeeman Hamiltonian are now
\begin{equation}
 \left<mps \left|H_Z \right|m'p's' \right> = -\frac{1}{2}g_{\rm eff}\mu_B B\delta_{mm'} \delta_{pp'} 
\delta_{-s,s'},
\nonumber
\end{equation}
and the SOI terms become
\begin{equation}
 \left<mps \left|H_R \right|m'p's' \right> = \left\{ 
  \begin{array}{l l}
   \vspace{1mm}  -\frac{\alpha}{2R}ms\delta_{mm'}\delta_{ss'} + \\
   \vspace{1mm} \frac{\alpha R}{4\ell_B^2} \left[ (1-s)\delta_{m,m'+2} - \delta_{m,m'} \right] \delta_{-s,s'}  \\ 
+ {\rm H.\ c.} \hspace{5mm} \text{if $p=p',$}\\
\\ \\
    2\frac{\alpha}{L}\frac{pp'}{p^2 - p'^2}(1-s)\delta_{m,m'+1}\delta_{-s,s'} \\ 
   \vspace{1mm} + {\rm H.\ c.} \hspace{5 mm} \text{if $p+p'$ odd,}\\
\\ \\
 0 \hspace{12 mm} \text{otherwise,}\\
  \end{array} \right.
\nonumber
\end{equation}
\\

\begin{equation}
 \left<mps \left|H_D \right|m'p's' \right> = \left\{ 
  \begin{array}{l l}
    \vspace{1mm} i \frac{\beta}{4 R} (1-s)(2m-1) )\delta_{m,m'+1} \delta_{-s,s'} + \\ 
    \vspace{1mm} i\frac{\beta R}{2\ell_B^2} \delta_{m,m'+1} s\delta_{ss'} \\ 
    + {\rm H.\ c.} \hspace{5 mm} \text{if $p=p',$}  \\
\\ \\
    2i\frac{\beta}{L}\frac{pp'}{p^2 - p'^2}s\delta_{mm'}\delta_{ss'} \\ 
  \vspace{1mm} + {\rm H.\ c.} \hspace{5 mm} \text{if $p+p'$ odd,} \\
\\ \\
 0 \hspace{12 mm} \text{otherwise.}\\
  \end{array} \right.
\nonumber
\end{equation}


\begin{thebibliography}{}

\bibitem{McCord1990}
M. A McCord and D. D. Awschalom, Appl. Phys. Lett. {\bf 57}, 2153 (1990)

\bibitem{Peeters1993}
F. M. Peeters and A. Matulis, Phys. Rev. B {\bf 48}, 15166 (1993)

\bibitem{Matulis1994}
A. A. Matulis, F. M. Peeters, and P. Vasilopoulos, 
Phys. Rev. Lett., {\bf 72 }, 1518 (1994)
\red{ \bibitem{Cerchez2007}
M. Cerchez, S. Hugger, T. Heinzel, and N. Schulz,
Phys. Rev. B  {\bf 75}, 035341 (2007).
} ̈
\red{ \bibitem{Tarasov2010}
A. Tarasov, S. Hugger, H. Xu, M. Cerchez, T. Heinzel, I. V. Zozoulenko,
U. Gasser-Szerer, D. Reuter, and A. D. Wieck,
Phys. Rev. Lett., {\bf 104}, 186801 (2010).
}

\bibitem{Carmona1995}
H. A. Carmona, A. K. Geim, A. Nogaret, P. C. Main, J. T Foster, M. Henini, 
S. P. Beaumont, and M. G. Blamire, Phys. Rev. Lett.  {\bf 74}, 3009 (1995)

\bibitem{Ye1995}
P. D. Ye, D. Weiss, R. R. Gerhardts, M. Seeger, M.  K. von Klitzing,
K. Eberl, and H. Nickel, Phys. Rev. Lett.  {\bf 74}, 3013 (1995)

\bibitem{Muller1992}
J. E. M\"uller, Phys. Rev. Lett. {\bf 68}, 385 (1992)

\bibitem{Ibrahim1995}
S. I. Ibrahim and F. M. Peeters, Phys. Rev. B {\bf 52}, 17321 (1995)

\bibitem{Zwerschke1999}
S. D. M. Zwerschke, A. Manolescu, and R. R. Gerhardts,
Phys. Rev. B {\bf 60}, 5536 (1999)

\bibitem{Cho2006}
A. Cho, Science {\bf 313}, 164 (2006)

\bibitem{Friedland2007}
K. -J. Friedland, R. Hey, H. Kostial, A. Riedel, and K. H. Ploog,
Phys. Rev. B {\bf 75}, 045347 (2007)

\bibitem{Friedland2009}
K. -J. Friedland, A. Siddiki, R. Hey, H. Kostial, and A. Riedel,
Phys. Rev. B {\bf 79}, 125320 (2009)

\bibitem{Gayer2012}
X. X. Gayer, X. X.,
DPG Meeting, Berlin, March 2012.

\bibitem{Friedland2008}
K. -J Friedland, R. Hey, H. Kostial, and A. Riedel,
phys. stat. sol. (c) {\bf 5}, 2850 (2008)

\bibitem{Thelander2006}
C. Thelander, P. Agarwal, S. Brongersma, J. Eymery, L. Feiner, 
A. Forchel, M. Scheffler, W. Riess, B. Ohlsson, U. G\"osele, L. and Samuelson,
Materials Today, {\bf 9} 28 (2006)

\bibitem{Bryllert2006}
T. Bryllert, L. E. Wernersson, L. E. Fröberg, and L. Samuelson,
IEEE Electron Device Letters {\bf 27}, 323 (2006)

\bibitem{Richter2008}
T. Richter, C. Bl\"omers, H. L\"uth, R. Callarco, M. Indlekoffer, M. Marso, T Sch\"apers,
Nano Lett. {\bf 8}, 2834 (2008)

\bibitem{Blomers2008}
C. Bl\"omers, T. Sch\"apers, T. Richter, R. Callarco, H. L\"uth, and M. Marso,
Phys. Rev. B {\bf 77}, 201301(R) (2008)

\bibitem{Wirths2011}
S. Wirths, K. Weis, A. Winden, K. Sladek, C. Volk, S. Alagha, T. E. Weirich, M. von der Ahe, 
H. Hardtdegen, H. L\"uth, N. Demarina, D. Gr\"utzmacher, and T. Sch\"apers,
Journal of Applied Physics {\bf 110}, 053709 (2011)

\bibitem{Blomers2011}
C. Bl\"omers, M. I. Lepsa, M. Luysberg, D. Gr\"utzmacher, H. L\"uth H. and T. Sch\"apers,
Nano Lett. {\bf 11}, 3550 (2011)

\bibitem{Blomers2013}
C. Bl\"omers, T. Rieger, P. Zellekens, F. Haas, 
  M. I. Lepsa, H. Hardtdegen, \"O. G\"ul, N. Demarina, D. Gr\"utzmacher, H. L\"uth, 
  and T. Sch\"apers,
Nanotechnology {\bf 24}, 035203 (2013)

\bibitem{Blomers2012}
C. Bl\"omers, T. Grap, M. I. Lepsa, J. Moers, S. Trellenkamp, D. Gr\"utzmacher, 
H. L\"uth and T. Sch\"apers, Appl. Phys. Lett. {\bf 101}, 152106 (2012)

\bibitem{Tserkovnyak2006}
Y. Tserkovnyak and B. I. Halperin Phys. Rev. B {\bf 74},  245327 (2006)

\bibitem{Gladilin2013}
V. N. Gladilin, J. Tempere, J. T. Devreese, and P. M. Koenraad,
Phys. Rev. B {\bf 87},  165424 (2013)

\bibitem{Ferrari2008}
G. Ferrari, A. Bertoni, G. Goldoni, and E. Molinari,
Phys. Rev. B {\bf 78},  115326 (2008)

\bibitem{Ferrari2009a}
G. Ferrari, G. Cuoghi, A. Bertoni, G. Goldoni, and E. Molinari,
J.Phys.: Conf. Ser. {\bf 193}, 012027 (2009)

\bibitem{Ferrari2009b}
G. Ferrari, G. Goldoni, A. Bertoni, G. Cuoghi, and E. Molinari,
Nano Lett. {\bf 9}, 1631 (2009)

\bibitem{Royo2013}
M. Royo, A. Bertoni, and G. Goldoni,
Phys. Rev. B {\bf 87}, 115316  (2013)

\bibitem{Wong2011}
B. M. Wong, F. L\'eonard, Q. Li, and T. Wang, T.,
Nano Lett., {\bf 11}, 3074 (2011)

\bibitem{Rieger2012}
T. Rieger, M. Luysberg, T. Sch\"apers, D. and Gr\"tzmacher, D. and
M. I. Lepsa,
Nano Lett. {\bf 12}, 5559, (2012)

\bibitem{Haas2013}
F. Haas, K. Sladek, A. Winden, M. von der Ahe, T. E. Weirich, T. Rieger,
H. and L\"uth, D. Gr\"tzmacher, T. Sch\"apers, and H. Hardtdegen,
Nanotechnology {\bf 24}, 085603 (2013)

\bibitem{Bellucci2010}
S. Bellucci and P. Onorato, Phys. Rev. B {\bf 82}, 205305 (2010).

\bibitem{Bringer2011}
A. Bringer and T. Sch\"apers, Phys. Rev. B {\bf 83}, 115305 (2011)

\bibitem{Winkler}
R. Winkler,
{\em Spin orbit coupling effects in two-dimensional electron and hole systems},
Springer-Verlag Berlin, Heidelberg, New York (2003)

\bibitem{Ihn}
T. Ihn,
{\em Semiconductor nanostructures. Quantum states and electronic transport.},
Oxford University Press (2010)

\bibitem{Fasth2007}
C. Fasth, A. Fuhrer, L. Samuelson, V. N. Golovach, and D. Loss,
Phys. Rev. Lett. {\bf 98}, 266801 (2007)

\bibitem{Joyce2010}
H. J. Joyce, J. Wong-Leung, Q. Gao, H. H. Tan, and C. Jagadish C.,
Nano Lett. {\bf 10}, 908 (2010)

\bibitem{Vleck1932}
J. H. van Vleck,
{\em The theory of Electric and Magnetic  Susceptibilities },
Oxford University Press, London (1932)
\red{ \bibitem{Moroz2000}
A. V. Moroz and C. H. W. Barnes,
Phys. Rev. B  {\bf 61}, R2464 (2000); 
Phys. Rev. B  {\bf 60}, 14272 (1999)}  

\bibitem{Sheng2006}
J. S. Sheng and K. Chang,
Phys. Rev. B  {\bf 74}, 235315 (2006)

\bibitem{Nowak2009}
J. S. Nowak and B. Szafran
Phys. Rev. B {\bf 80}, 195319 (2009)

\bibitem{Daday2011}
C. Daday, A. Manolescu, D. C. Marinescu, and V. Gudmundsson,
Phys. Rev. B {\bf 84}, 115311 (2011)

\bibitem{Serra2007}
L. Serra, D. Sanchez and R. Lopez, Phys.\ Rev.\ B {\bf 76}, 045339 (2007).
\red{
\bibitem{Bellucci2003}
S. Bellucci and P. Onorato, Phys. Rev. B {\bf 68}, 245322 (2003).}

\end{thebibliography}
\end{document}